# Automatic Detection of Performance Anomalies in Task-Parallel Programs

Andi Drebes, Karine Heydemann and Nathalie Drach
Sorbonne Universités, UPMC Univ Paris 06, CNRS
Laboratoire d'Informatique de Paris 6 (LIP6), UMR 7606
4, Place Jussieu
F-75005 Paris, France

Antoniu Pop
The University of Manchester
School of Computer Science
Oxford Road, Manchester M13 9PL
United Kingdom

Albert Cohen
INRIA and École Normale Supérieure
Département d'Informatique (DI), UMR 8548
45, Rue d'Ulm
F-75005 Paris, France

## I. Introduction

To efficiently exploit the resources of new many-core architectures, integrating dozens or even hundreds of cores per chip, parallel programming models have evolved to expose massive amounts of parallelism, often in the form of fine-grained tasks. Task-parallel languages, such as OpenStream [7], X10 [2], Habanero Java and C [1] or StarSs [6], simplify the development of applications for new architectures, but tuning task-parallel applications remains a major challenge. Performance bottlenecks can occur at any level of the implementation, from the algorithmic level (e.g., lack of parallelism or over-synchronization), to interactions with the operating and run-time systems (e.g., data placement on NUMA architectures), to inefficient use of the hardware (e.g., frequent cache misses or misaligned memory accesses); detecting such issues and determining the exact cause is a difficult task.

In previous work, we developed Aftermath [3], an interactive tool for trace-based performance analysis and debugging of task-parallel programs and run-time systems. In contrast to other trace-based analysis tools, such as Paraver [5] or Vampir [4], Aftermath offers native support for tasks, i.e., visualization, statistics and analysis tools adapted for performance debugging at task granularity. However, the tool currently does not provide support for the automatic detection of performance bottlenecks and it is up to the user to investigate the relevant aspects of program execution by focusing the inspection on specific slices of a trace file. In this paper, we present ongoing work on two extensions that guide the user through this process:

- a threshold-based analysis providing a high-level overview on a performance anomaly;
- and a linear regression approach correlating performance indicators and task duration to detect bottlenecks and match them with task-parallel execution.

## II. Aftermath

Aftermath operates on trace files generated during the execution of a task-parallel program and allows for an off-line analysis of dynamic events (after program termination). The main features can be summarized as follows:

- A visual representation of dynamic events, topological and temporal information (e.g., different run-time states a processor traverses over time, evolution of hardware performance counter values over time, task creation events, task synchronization and communication, memory accesses).
- Statistical views summarizing the data shown in the graphical representation and providing accurate numerical data (e.g., average task duration, a task duration distribution histogram, an incidence matrix summarizing communication between NUMA nodes).
- A detailed textual view with accurate information about a specific task instance selected by the user (e.g., time of creation, exact duration, data accesses).
- A powerful filtering interface allowing to limit the visualized information and statistics to precise aspects (e.g., specific processors, tasks of a certain duration or an application-specific type, tasks communicating with specific NUMA nodes).
- Creation of derived metrics (e.g., memory controller contention derived from data placement and data accesses, combining existing hardware performance counters, parallelism on average per interval).

Currently, the user has to detect bottlenecks manually, i.e., by selecting appropriate views and by applying the filters that highlight a performance anomaly and emphasize its cause. For example, if the user suspects that low performance is due to a high cache miss rate of a certain type of tasks (e.g., tasks carrying out matrix multiplications), he needs to visualize the task duration, limit the view to tasks of that type, inspect the cache miss rate of slow and fast tasks and develop the hypothesis of a correlation between task duration and the cache miss rate. The user must then repeat these steps for every potential source of performance anomalies.

## III. High-level analysis based on thresholds

Aftermath traces contain information on the different states each processor traverses over time. For example, a processor might create a task, entering the state associated to task creation, then execute this task in the associated execution state, and finally spend some time in a work-stealing state where it searches for new tasks to execute. Inspecting these states can help to establish whether a bottleneck arises from the run-time system or the application and thus give a first, high-level overview on a performance anomaly.

Ideally, each processor effectively contributes to the overall computation and spends nearly all of its time in the task execution state. Assume there are $n$ processors and an interval of duration $d$. Let $t_e$ be a threshold defining the target execution time in the task execution state; e.g., $t_e = 0.95$ indicates that at least 95% of the time should be spent in that state.

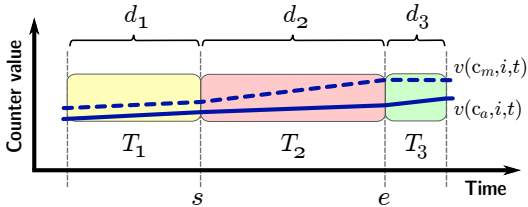

Fig. 1. Example of three tasks $T_1, T_2$ and $T_3$ of different durations $d_1, d_2$ and $d_3$ and the evolution of two hardware counter $c_a$ and $c_m$ counting the number of cache accesses and cache misses, respectively.

Let further $d_{e,i}$ be the overall duration a processor $i$ spends in task execution state within the interval. If the inequality $\sum_{i=1}^{n} d_{e,i} < t_e \cdot n \cdot d$ holds, there is not enough parallelism to saturate the machine with running tasks.

The root cause for insufficient parallelism can be refined with further threshold-based analysis. For example, if more than a fraction $t_c$ is spent in the task creation state, task creation overhead is likely to be too high or if a fraction of time $t_s$ is spent on work-stealing, then there might simply not be enough parallelism exposed by the application or there might be a load-balancing problem.

## IV. Correlating performance indicators with task durations

The analysis presented above does not cover performance anomalies that occur during the execution of tasks, such as in the cache miss rate scenario mentioned earlier. Hardware performance counters can provide insight on what happens during task execution. Aftermath already provides a generic interface for analysis of arbitrary performance counters, but it is up to the parallel run-time or the instrumented application to select the appropriate event type, to capture counter samples and to write them to the trace file. Manually determining the relevant event types can be time-consuming due to the multitude of events that can be monitored on modern architectures. Hence, automating this process is highly desirable. We propose a method based on linear regression in order to determine automatically whether a specific hardware event should be considered for performance analysis and under which conditions it is relevant.

Hardware performance counter values are usually captured per processor and first need to be associated with task instances. Assume that $v(c, i, t)$ returns the absolute value of a counter $c$ for processor $i$ at a time $t$. Determining how much a counter has changed during execution of a task is done by comparing $v$ at the beginning $s$ and the end $e$ of the task instance. Figure 1 illustrates this for two counters $c_a$ and $c_m$ tracking cache accesses and cache misses, respectively. The number of accesses generated by task $T_2$, is $N_{a,T_2} = v(c_a, i, e) - v(c_a, i, s)$ and the number of misses is $N_{m,T_2} = v(c_m, i, e) - v(c_m, i, s)$. Hence, the cache miss rate of $T_2$, is $R_{T_2}$ with:

$$R_{T_2} = \frac{N_{m,T_2}}{N_{a,T_2}} = \frac{v(c_m, i, e) - v(c_m, i, s)}{v(c_a, i, e) - v(c_a, i, s)}$$

The miss rates of $T_1$ and $T_3$, $R_{T_1}$ and $R_{T_3}$ can be determined similarly.

We consider that a performance indicator is relevant for performance analysis if the variation of task durations is sufficiently high—based on confidence intervals—and if the duration correlates with the performance indicator according to a linear model. In Figure 1, durations $d_i$ differ substantially and depend on the miss rates $R_{T_i}$ with $d_i \approx \alpha \cdot R_{T_i} + \beta$, $\alpha$ and $\beta$ being constant. Such relationships can easily be detected automatically by performing linear regressions and by comparing the coefficients of determination to a threshold value. However, they are rarely valid for all tasks. Hence, it is necessary to group task instances according to specific attributes, such as the task type or the executing processor, and to repeat the whole process for each group. At the end of the analysis, the user is informed which indicators are relevant and for which task types and processors this is the case.

## V. Conclusion and Outlook

We presented two simple automated techniques based on thresholds and linear regression that point the user more quickly to the relevant parts of a trace file and indicate which performance indicators should be taken into consideration for a detailed manual analysis. These have proven invaluable when performed manually [3] and are currently being implemented in Aftermath, a tool for the performance analysis and debugging of task-parallel applications.

In future work, we will take into account dynamic information on task and data placement and explore how more elaborated approaches, such as machine learning can be applied. Until now, we have implicitly assumed that the amount of work per task of a given type remains constant. This assumption might not hold in general, where longer tasks may exhibit higher values for a given performance indicator in the absence of a performance anomaly. To avoid such misleading indications, we will extend the analysis with an estimation of the amount of work performed by each task.